\documentclass[preprint,12pt]{elsarticle}

 \usepackage{subcaption}

\usepackage{amssymb}
\usepackage{amsmath}
\usepackage{cleveref}
 \usepackage{lineno}

\journal{Nuclear Physics B}

\begin{document}

\begin{frontmatter}



\title{Triplet assembly and certification of the new generation of RPC for the ATLAS phase-2 upgrade at Max Planck Institute}


\author{G.Proto on behalf of the ATLAS Muon Community} 
\affiliation{organization={Max Planck Institut fur Physik},
            addressline={Boltzmannstr. 8}, 
            city={Munich},
            postcode={85748}, 
            state={Germany}}

\begin{abstract}
A new generation of Resistive Plate Chambers have been developed for the ATLAS phase-2 upgrade in sight of the High-Luminosity phase of the Large Hadron Collider. These RPCs consist of three independent 1 mm gas gaps(singlets) equipped with a newly low-threshold Front-End electronics, assembled in the same mechanical structure (triplet). During 2024 the production of the phase-2 RPCs started and the detectors will undergo (2024-2025) a certification test before the installation in the ATLAS cavern. The triplet assembly and the certification with cosmic rays of the BIS-type detectors is performed at the Max-Planck-Institute for Physics (MPI) in Munich. The architecture of the cosmic rays test stand has been built at MPI and has been studied in order to provide an efficient and robust structure to ensure an excellent quality control and study precisely the whole RPC performance needed to certify the detectors for the ATLAS experiment. In this presentation the assembly procedure, the architecture of the test stand and the certification protocols will be presented along with the validation tests to characterize the RPC performance.  
\end{abstract}

\begin{graphicalabstract}
\end{graphicalabstract}






\end{frontmatter}
\footnote{Copyright 2025 CERN for the benefit of the ATLAS Collaboration. Reproduction of this article or parts of it is allowed as specified in the CC-BY-4.0 license.}

\section{ATLAS RPC phase-2 upgrade }
\label{sec1}
\begin{figure}[!htbp]
\centering
\begin{subfigure}{0.45\textwidth} 
    \centering
    \includegraphics[width=\textwidth]{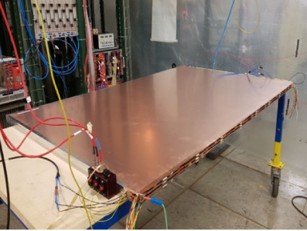}
    \caption{}
    \label{fig:singlets}
\end{subfigure}
\hfill 
\begin{subfigure}{0.45\textwidth} 
    \centering
    \includegraphics[width=\textwidth]{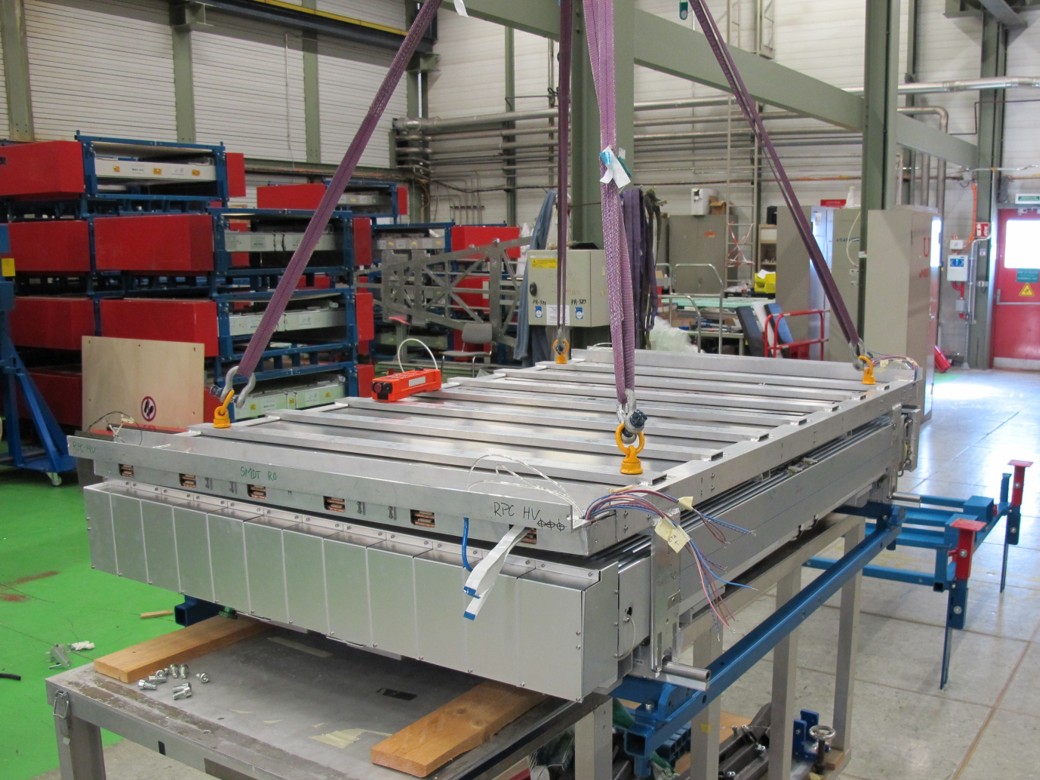}
    \caption{}
    \label{fig:triplet}
\end{subfigure}
\caption{Singlets (a) and RPC triplet integrated with s-MDT (b) layouts. \label{fig:Detectorlo}}
\end{figure}

In sight of the High-Luminosity Large Hadron Collider (HL-LHC) program, the ATLAS Muon Spectrometer (MS) must be able to operate at instantaneous luminosity of $L$ = $\rm{7.5 \times 10^{34}}$ $\rm{cm^{-2}s^{-1}}$. The detector system will undergo changes accordingly in order to cope with the higher flux of incident particles. In particular, the RPC system will be upgraded in order to provide an improvement of the acceptance, redundancy and selectivity of the trigger \cite{atlasdet}. The upgrade involves the installation of thin-gap RPCs (1 mm) in the Inner Barrel (BI) region of the ATLAS MS, where no RPC were present during previous LHC runs. These detectors are classified as a new generation of RPCs, due to their improved time resolution ($\sim$ 300 ps, enabled by the thinner gas gap) and new Front-End (FE) electronics capable of operating with a threshold of just a few fC \cite{feelectronics}. This design ensures reliable operation at rates of up to 500 Hz/$\rm{cm^{2}}$ throughout the experiment duration. \\The BI RPCs consist of three independent detectors, known as singlets (Figure \ref{fig:Detectorlo}\subref{fig:singlets}), embedded within the same mechanical structure to form a triplet (Figure \ref{fig:Detectorlo}\subref{fig:triplet}). Two layouts are implemented for the BI RPCs, depending on the ATLAS sector where they will be installed: BIL and BIS. A third layout is foreseen for BOR chambers, to be installed in 2 outer sectors (sectors 11 and 15).

\renewcommand\thesubfigure{\alph{subfigure}} 

The key distinction between the two lies in the integration of the BIS type with s-MDT detectors, resulting in a different mechanical structure designed to maintain electrical and mechanical independence between the two detector types.\\ The construction, certification, installation, and commissioning of the BI RPCs represent a collaborative effort among several institutes, including INFN, the Max Planck Institute for Physics (MPI), University of Science and Technology of China (USTC). The gas gaps are manufactured by General Tecnica in Italy and subsequently tested and assembled by their respective institutes—INFN \cite{mattia} for the BIL and USTC for the BIS. At this stage, singlets are created by integrating the gas gaps with the two strip panels equipped with the electronics. Following this, the singlets undergo preliminary certification test with cosmic rays before being integrated into the mechanical structure. Assembly is performed at CERN by the INFN group for the BIL type and in Munich by MPI for the BIS type. After assembly, the detectors are tested again to identify any potential damage caused during integration.\\Once certified, the detectors undergo additional tests at CERN before their installation and commissioning within the ATLAS experiment.
\section{Max Planck Institute responsibilities in the ATLAS phase-2 upgrade}
\label{sec2}
As part of the BI project, the MPI group is tasked with the assembly and certification of the BIS-type triplets, totaling 96 triplets (288 singlets). Additionally, MPI has designed the mechanical structure and the associated service boxes for these detectors and will play a key role in their integration with the s-MDT detectors. Furthermore, in collaboration with industrial partners, MPI will produce the RPC detectors to be installed in the BOR/BOM region (barrel outer region, sector 11 and 15) \cite{falla}. This paper focuses on the assembly and certification of the triplets, along with the plans for the integration with the s-MDT detectors.\\Once the production phase begins, MPI is supposed to receive 48 singlets under the responsibility of USTC people, and assemble 16 triplets.\\ The entire testing process for one detector consists in two phases : 
\begin{itemize}
    \item Standard checks of the singlets are performed alongside their assembly into the mechanical frame. This includes a visual inspection of the singlets, verification of gas tightness, testing of the electronics, and the characteristic HV-I curve. Following these checks, the detectors are mounted into the mechanical frame.
    \item The detectors are tested with cosmic rays. This stage focuses on evaluating detector noise and key performance parameters, including efficiency, cluster size, and time over threshold.
\end{itemize}

The entire procedure has been meticulously designed to efficiently meet the stringent requirements of the ATLAS experiment while ensuring the production of high-quality detectors for experimental use. 
\section{Triplet assembly }
\label{sec3}
\begin{figure}[!h]
\includegraphics[width=\textwidth]{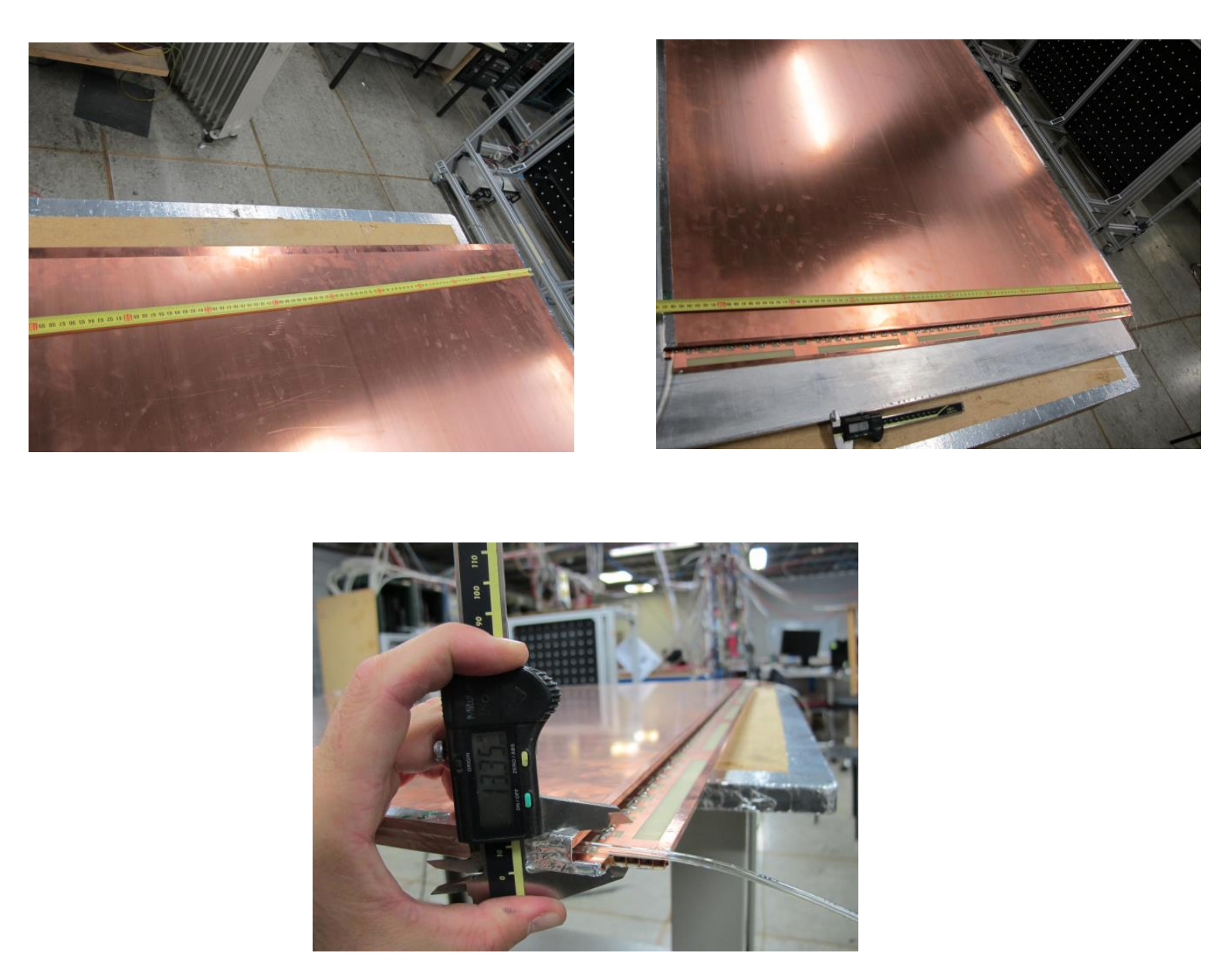}
\caption{Initial checks of the singlets:visual inspection and geometrical tests}\label{fig:checks}
\end{figure}

The triplet assembly phase begins with the unpacking of the singlets (gas gaps equipped with strip panels, electronics and services) and their sequential placement onto the designated table. General checks are conducted for each singlet to identify any significant issues that may have arisen during transportation, as illustrated in Figure \ref{fig:checks}.
This phase mainly involves a visual inspection of the singlets, with particular attention given to the FE electronics, as well as measurements of the outer dimensions. During this stage, the detectors are also flushed, and a quick data acquisition is performed to identify any dead channels, following the method described in Section \ref{sec2}.

\begin{figure}[!h]
\centering
\begin{subfigure}{0.45\textwidth} 
    \centering
    \includegraphics[width=\textwidth]{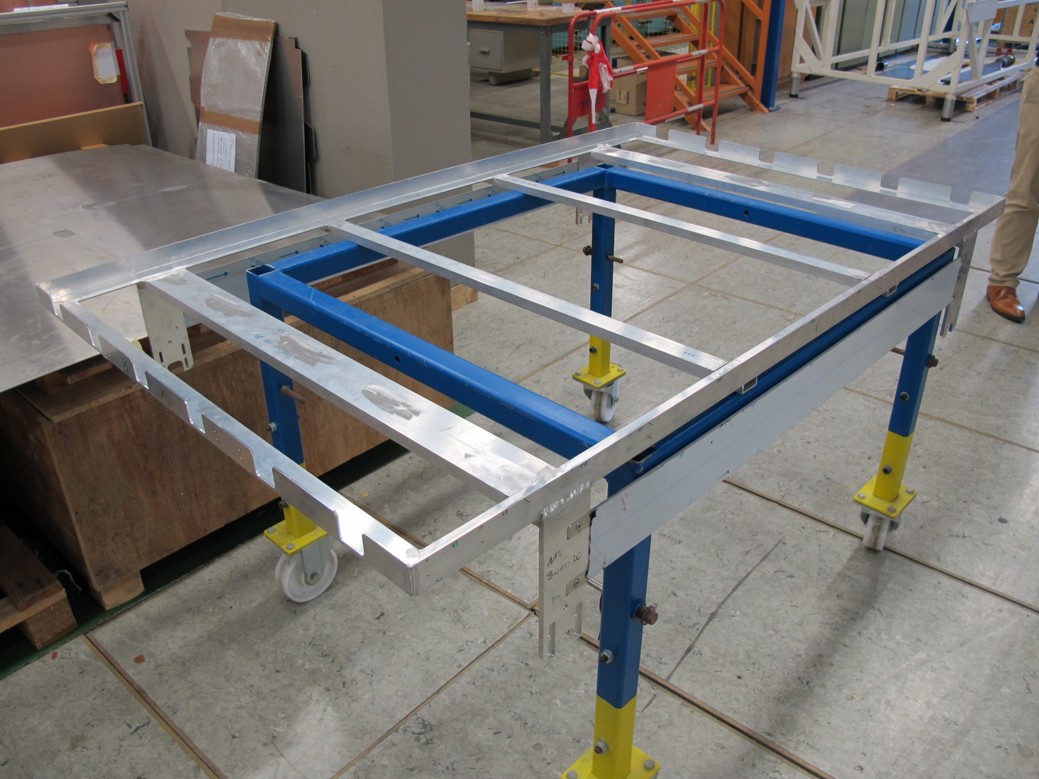}
    \caption{}
    \label{fig:mech1}
\end{subfigure}
\hfill 
\begin{subfigure}{0.45\textwidth} 
    \centering
    \includegraphics[width=\textwidth]{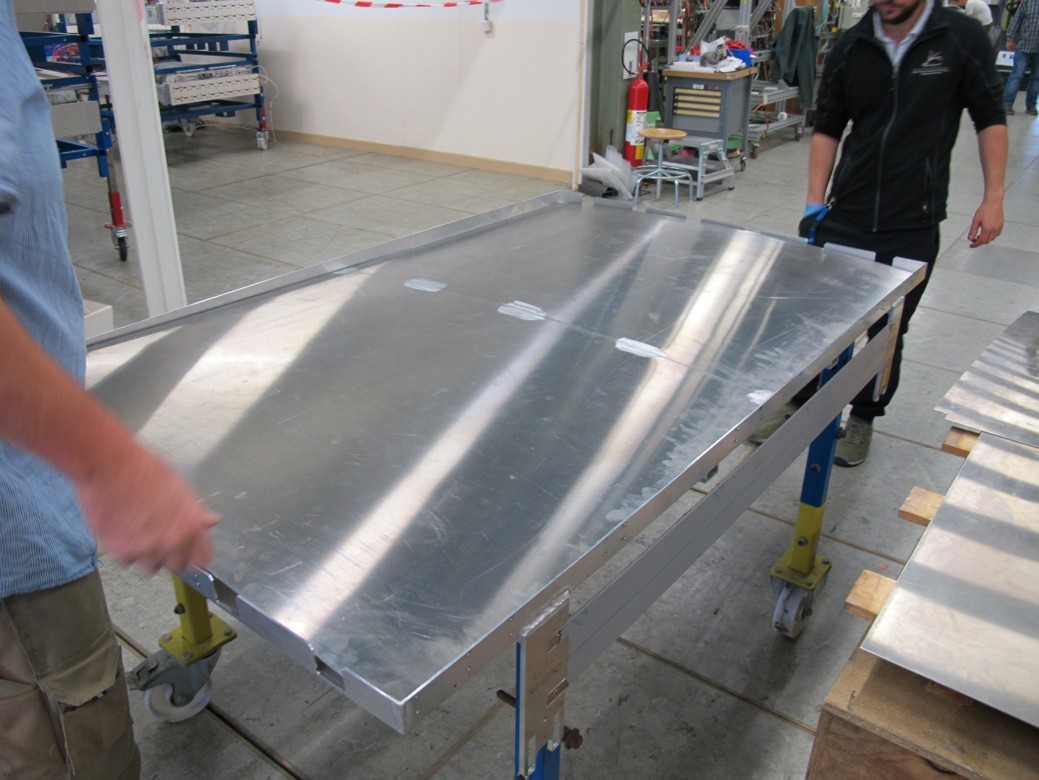}
    \caption{}
    \label{fig:mech2}
\end{subfigure}
\caption{Lower frame of the mechanical structure (a) and pre-bent plate (b).\label{fig:mechanics}}

\end{figure}

In parallel, the lower frame of the mechanical structure is positioned on a specially designed table, and the pre-bent plate is placed on top of it, as shown in Figure \ref{fig:mechanics}.
\renewcommand\thesubfigure{\alph{subfigure}} 

\begin{figure}[!h]
\centering
\begin{subfigure}{0.45\textwidth} 
    \centering
    \includegraphics[width=\textwidth]{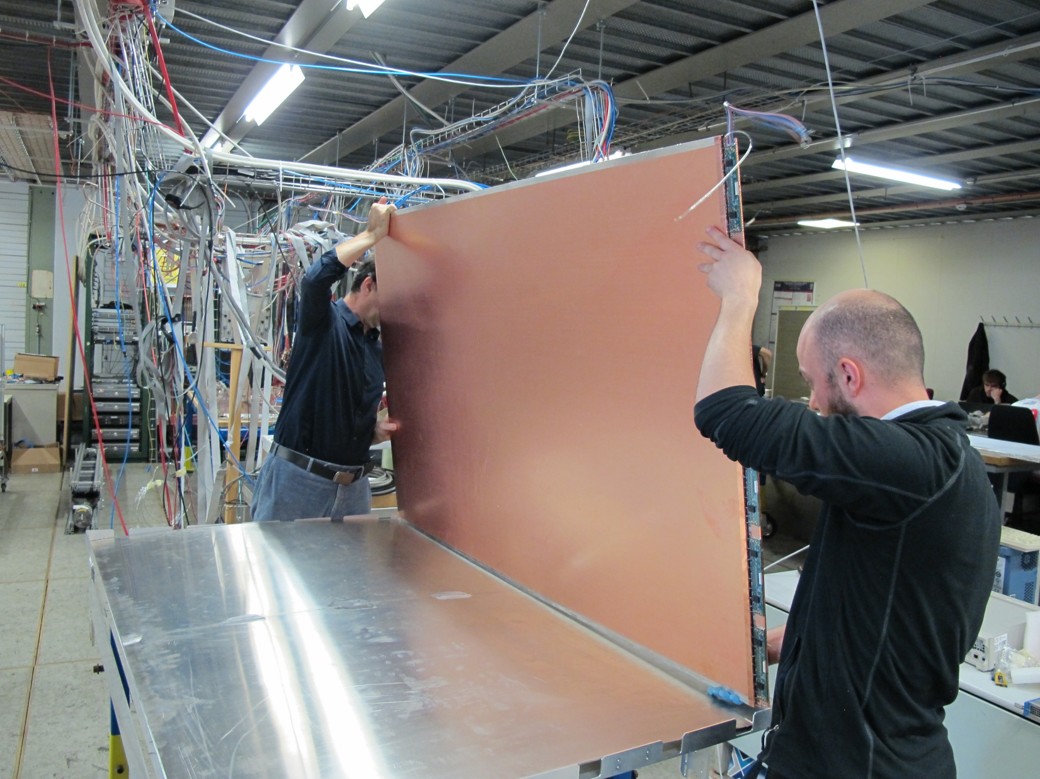}
    \caption{}
    \label{fig:mech1}
\end{subfigure}
\hfill 
\begin{subfigure}{0.45\textwidth} 
    \centering
    \includegraphics[width=\textwidth]{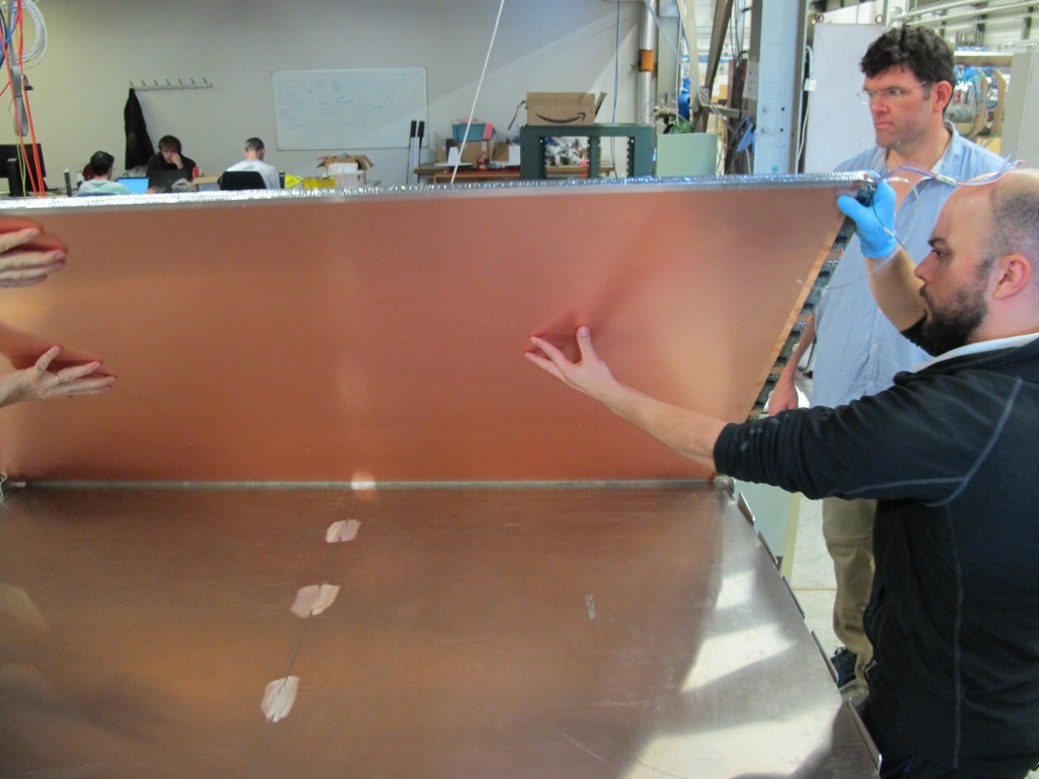}
    \caption{}
    \label{fig:mech2}
\end{subfigure}
\caption{Integration of the singlet within the mechanics \label{fig:integration}}
\end{figure}

Once the checks on the first singlet are succesfully completed, it is installed into the mechanical frame, following the procedure shown in Figure \ref{fig:integration}.
\renewcommand\thesubfigure{\alph{subfigure}} 

\begin{figure}[!h]
\centering
\begin{subfigure}{0.45\textwidth} 
    \centering
    \includegraphics[width=\textwidth]{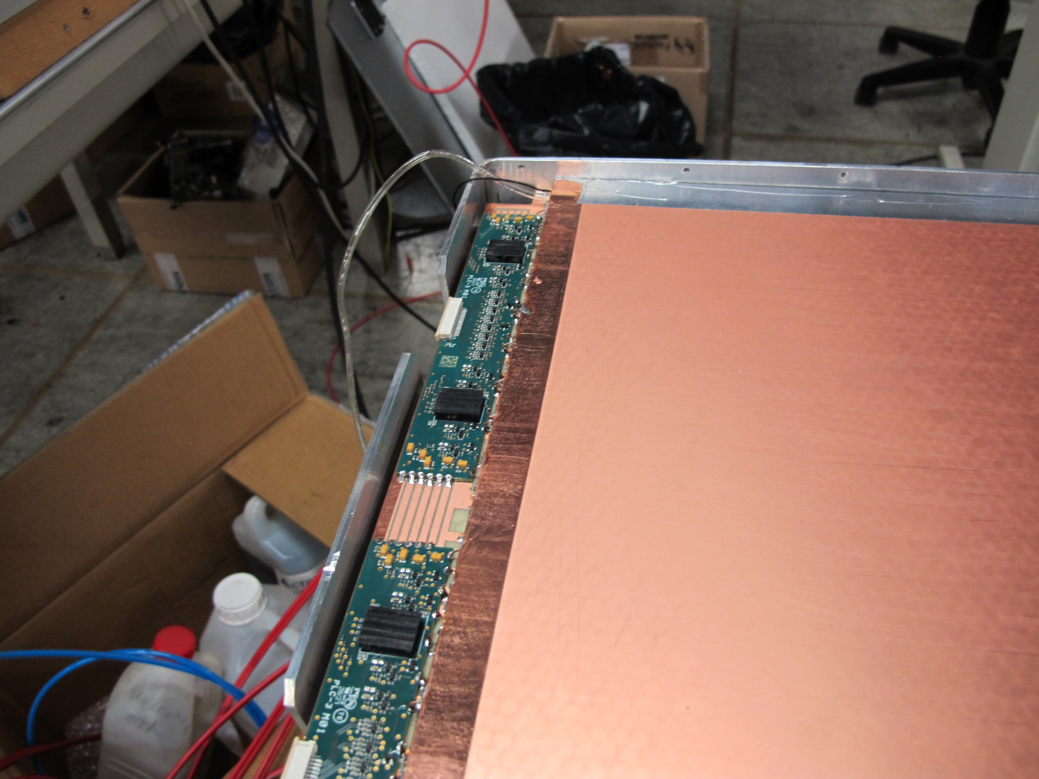}
    \caption{Check of the FE boards}
    \label{fig:acheck1}
\end{subfigure}
\hfill 
\begin{subfigure}{0.45\textwidth}
    \centering
    \includegraphics[width=\textwidth]{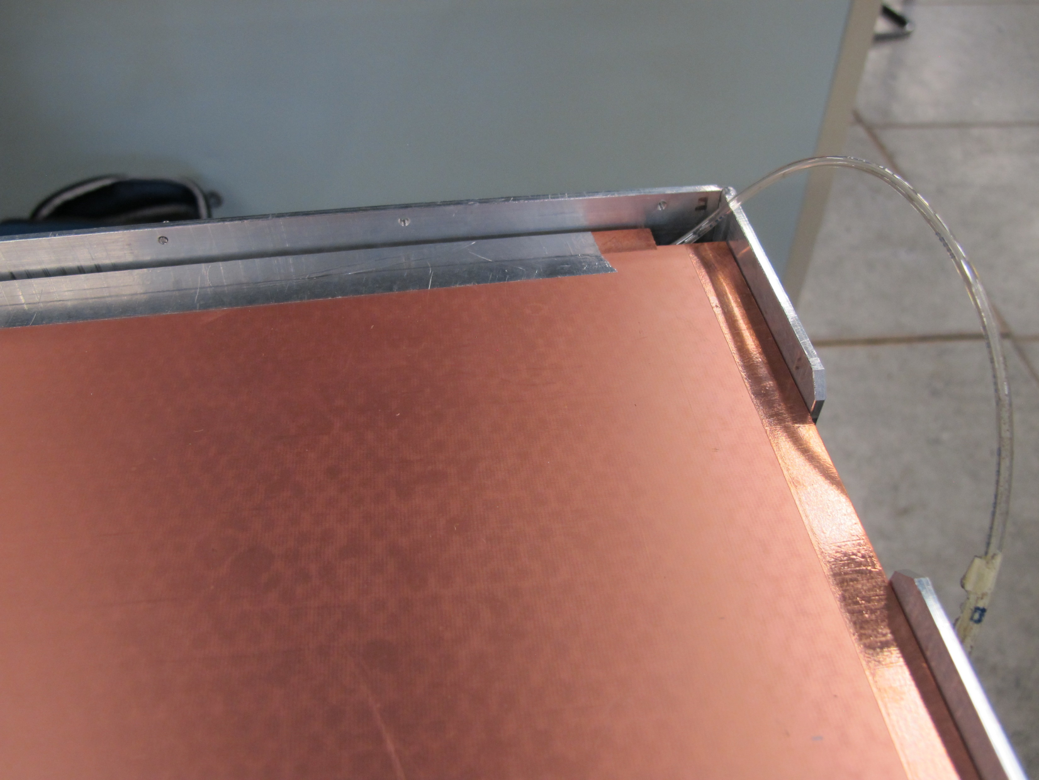}
    \caption{Gas pipe through cut-out}
    \label{fig:acheck22}
\end{subfigure}

\vspace{0.5cm} 

\begin{subfigure}{0.45\textwidth}
    \centering
    \includegraphics[width=\textwidth]{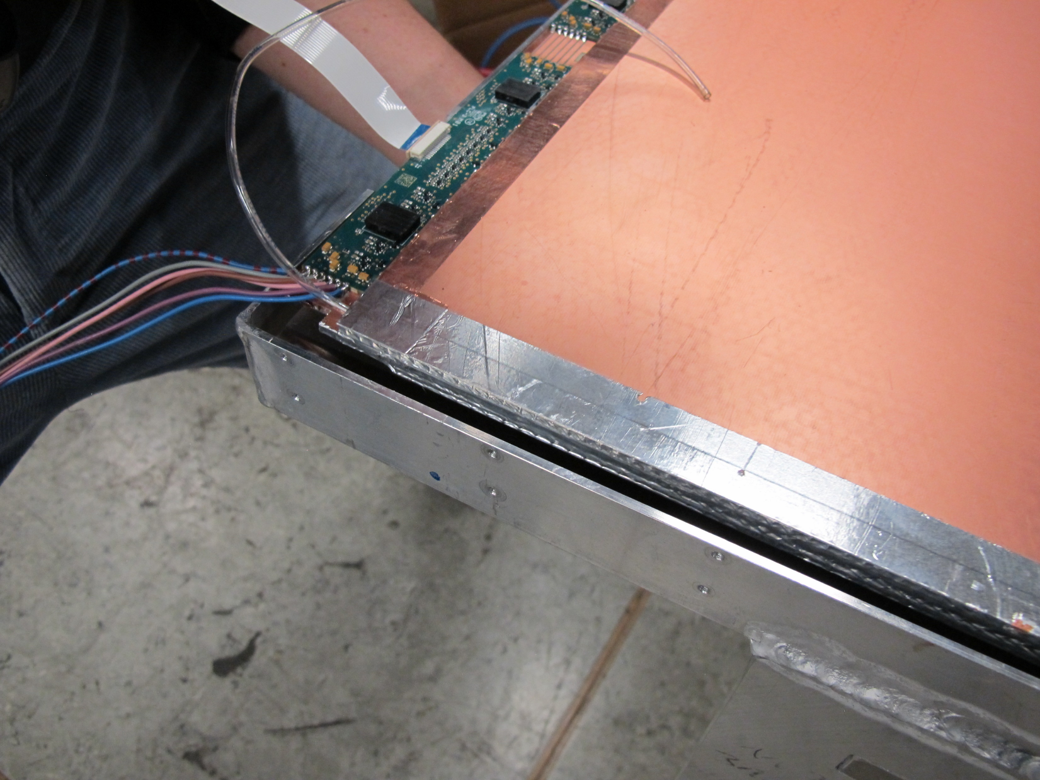}
    \caption{Low voltage through cut-out}
    \label{fig:acheck3}
\end{subfigure}
\hfill
\begin{subfigure}{0.45\textwidth}
    \centering
    \includegraphics[width=\textwidth]{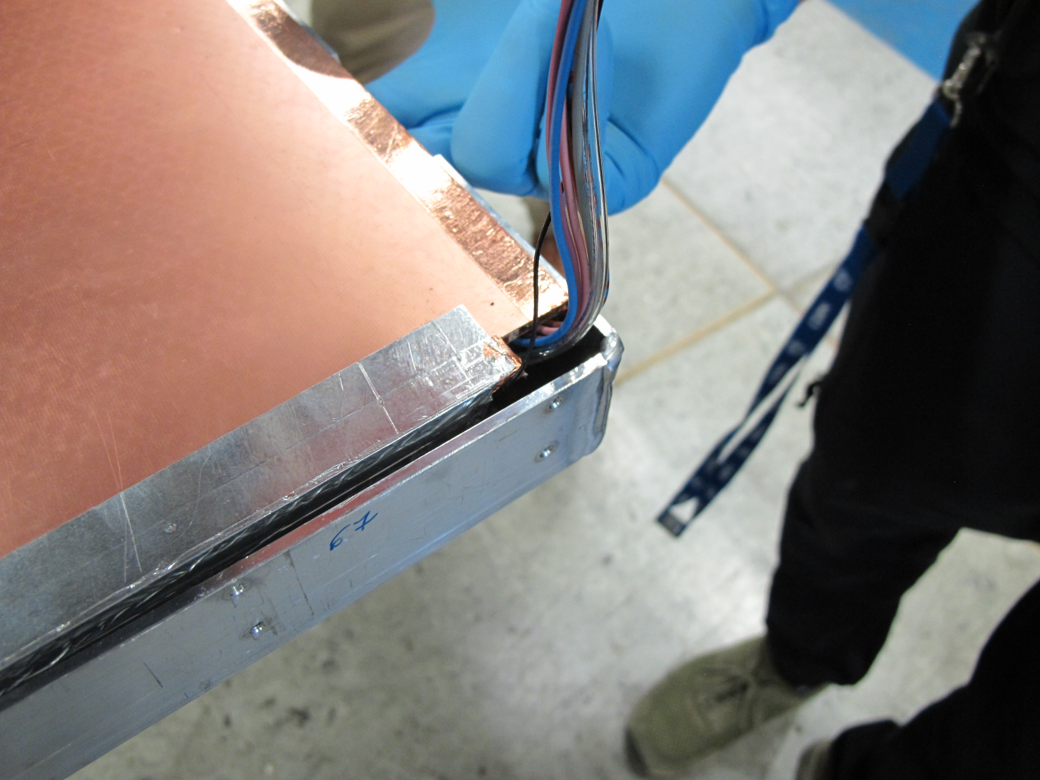}
    \caption{Low voltage opposite side through cut-out}
    \label{fig:acheck44}
\end{subfigure}

\caption{Checks performed on the singlet within the mechanics.\label{fig:checkinside}}

\end{figure}

This process allows to minimize the possibility to damage the FE electronics and services. After the first singlet is placed, additional checks are performed. These include verifying the correct routing of the services, such as gas pipes, low and high voltage lines, and taping, through the designated cut-outs, as shown in Figure \ref{fig:checkinside}. 

\begin{figure}[!h]
\centering
\begin{subfigure}{0.45\textwidth} 
    \centering
    \includegraphics[width=\textwidth]{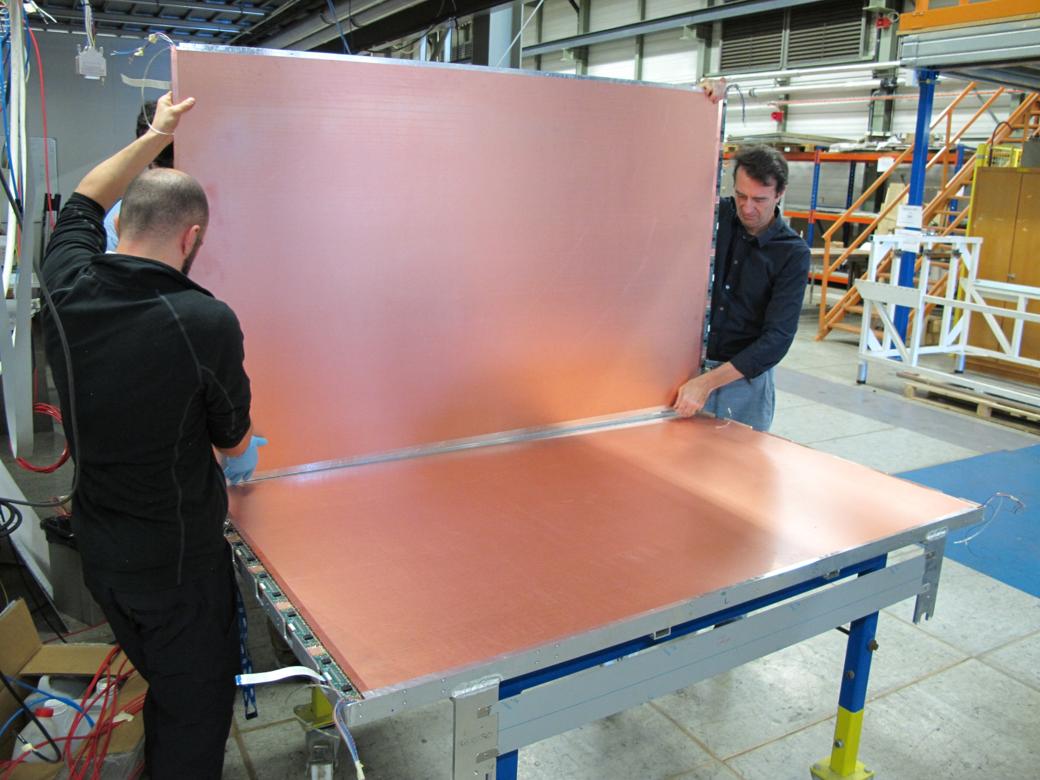}
    \caption{Positioning of the second singlet}
    \label{fig:ss1}
\end{subfigure}
\hfill 
\begin{subfigure}{0.45\textwidth}
    \centering
    \includegraphics[width=\textwidth]{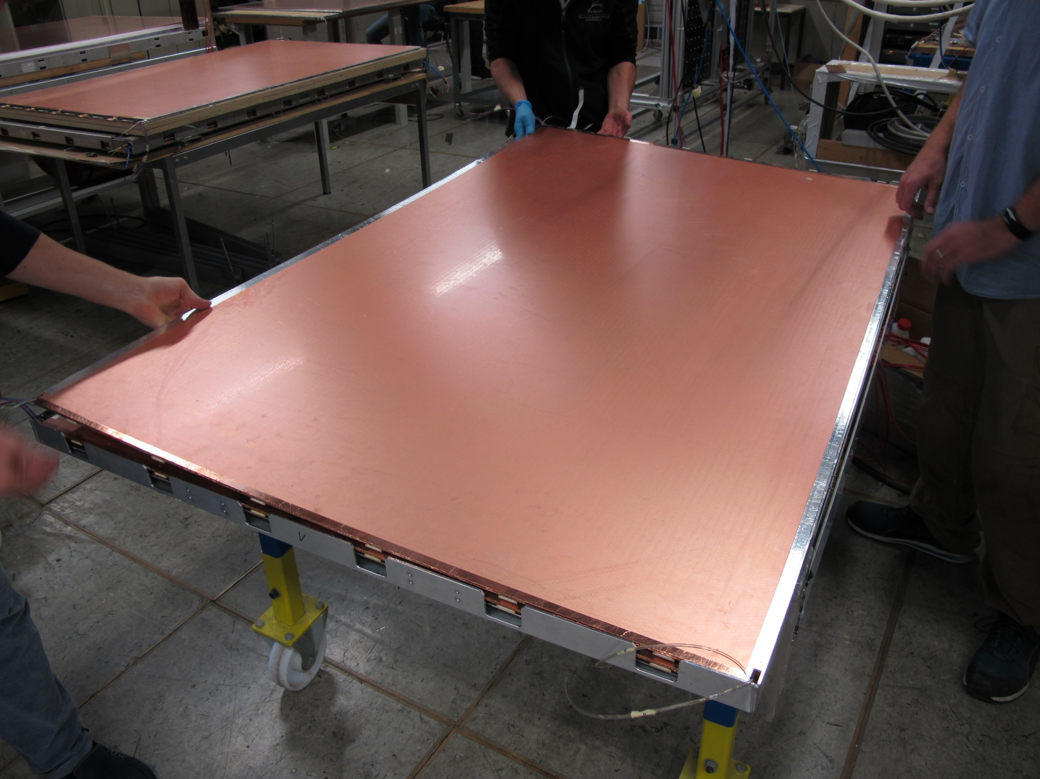}
    \caption{Second singlet within the mechanics}
    \label{fig:secondsing2}
\end{subfigure}

\vspace{0.5cm} 

\begin{subfigure}{0.45\textwidth}
    \centering
    \includegraphics[width=\textwidth]{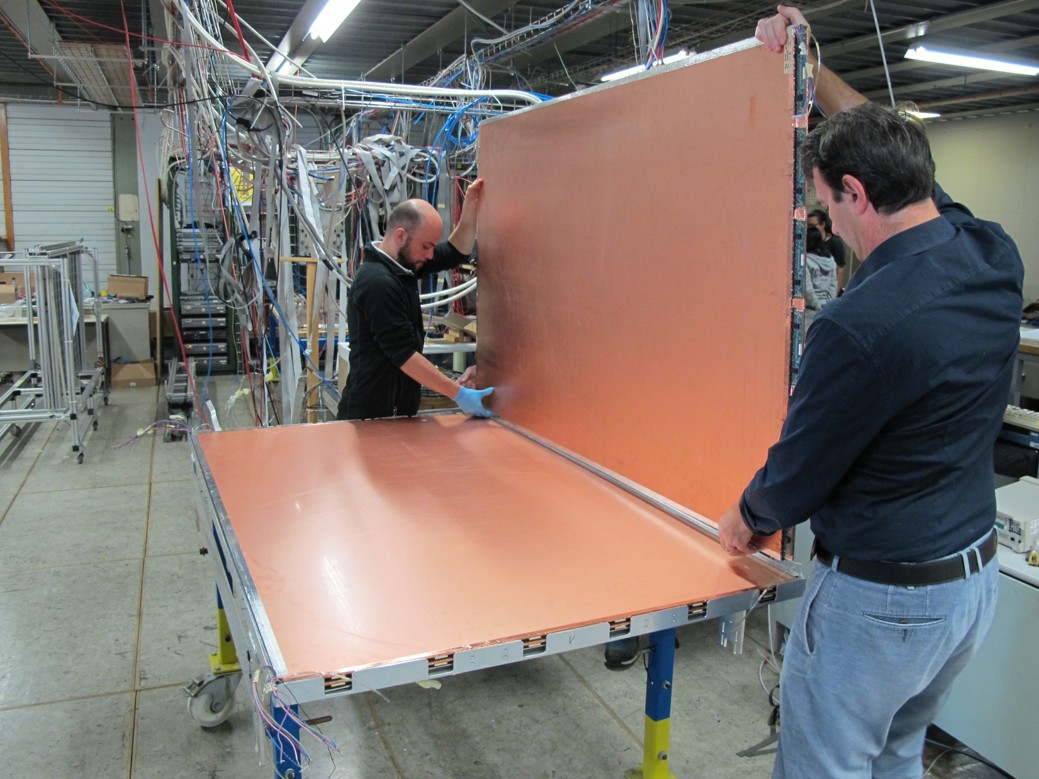}
    \caption{Positioning of the third singlet}
    \label{fig:thirdsing}
\end{subfigure}
\hfill
\begin{subfigure}{0.45\textwidth}
    \centering
    \includegraphics[width=\textwidth]{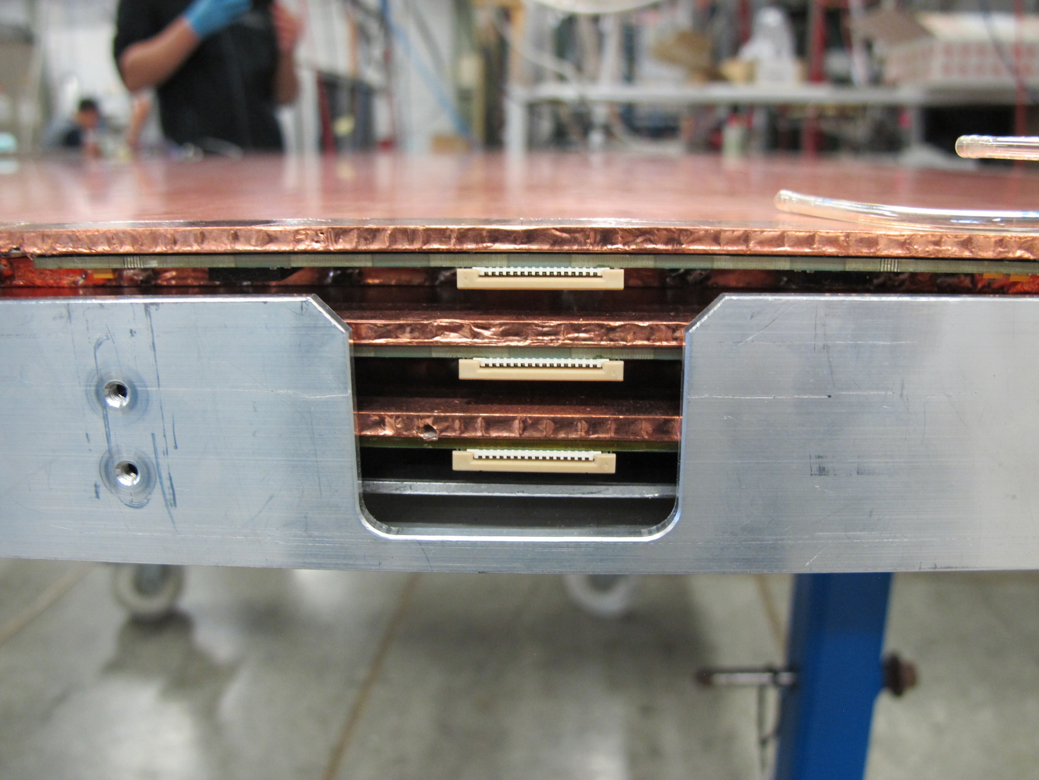}
    \caption{Three singlets installed and aligned}
    \label{fig:trip}
\end{subfigure}

\caption{Positioning of the singlets within the mechanics.\label{fig:tripcon}}

\end{figure}

The same procedure is repeated for the remaining two singlets, as in Figure \ref{fig:tripcon}. 

\begin{figure}[!htbp]
\centering
\begin{subfigure}{0.45\textwidth} 
    \centering
    \includegraphics[width=\textwidth]{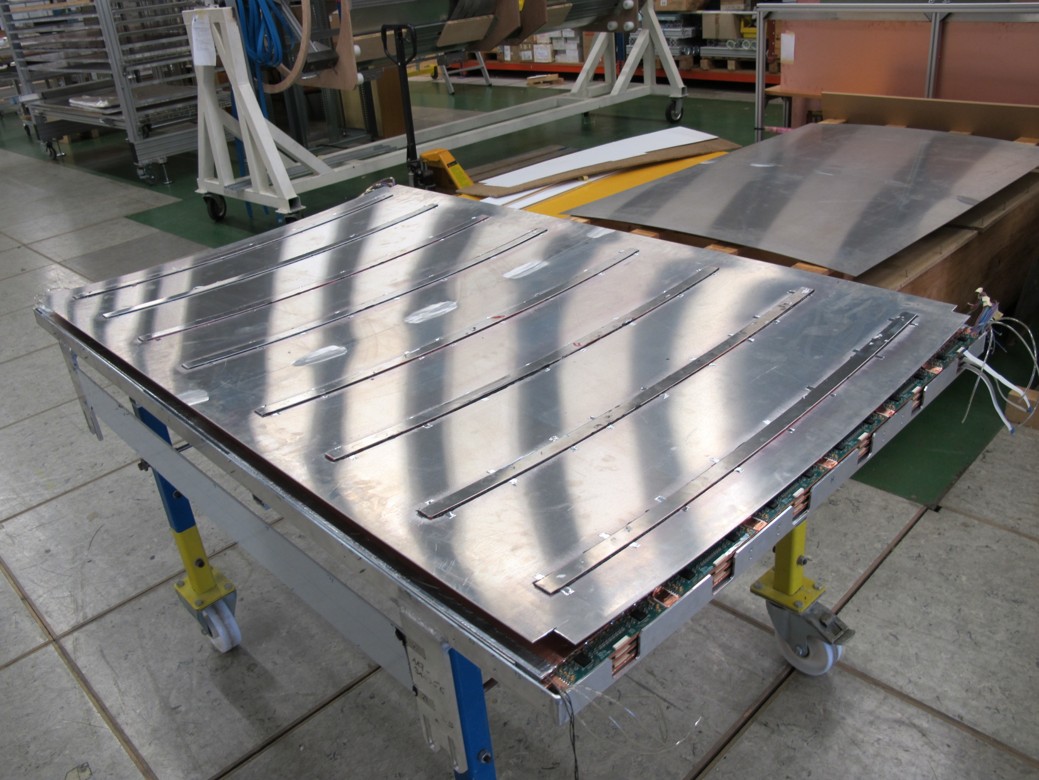}
    \caption{Positioning of the upper plate}
    \label{fig:upperplate}
\end{subfigure}
\hfill 
\begin{subfigure}{0.45\textwidth}
    \centering
    \includegraphics[width=\textwidth]{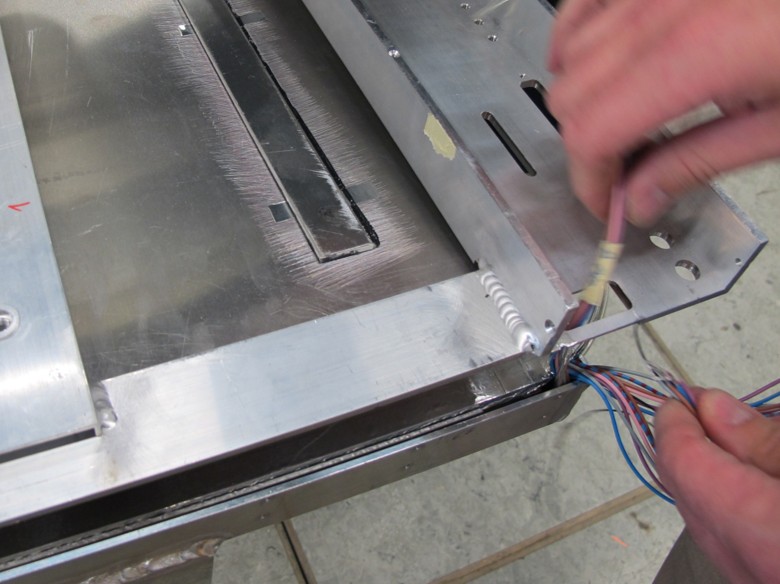}
    \caption{check services on cut-outs}
    \label{fig:mechcutout}
\end{subfigure}

\vspace{0.5cm} 

\begin{subfigure}{0.45\textwidth}
    \centering
    \includegraphics[width=\textwidth]{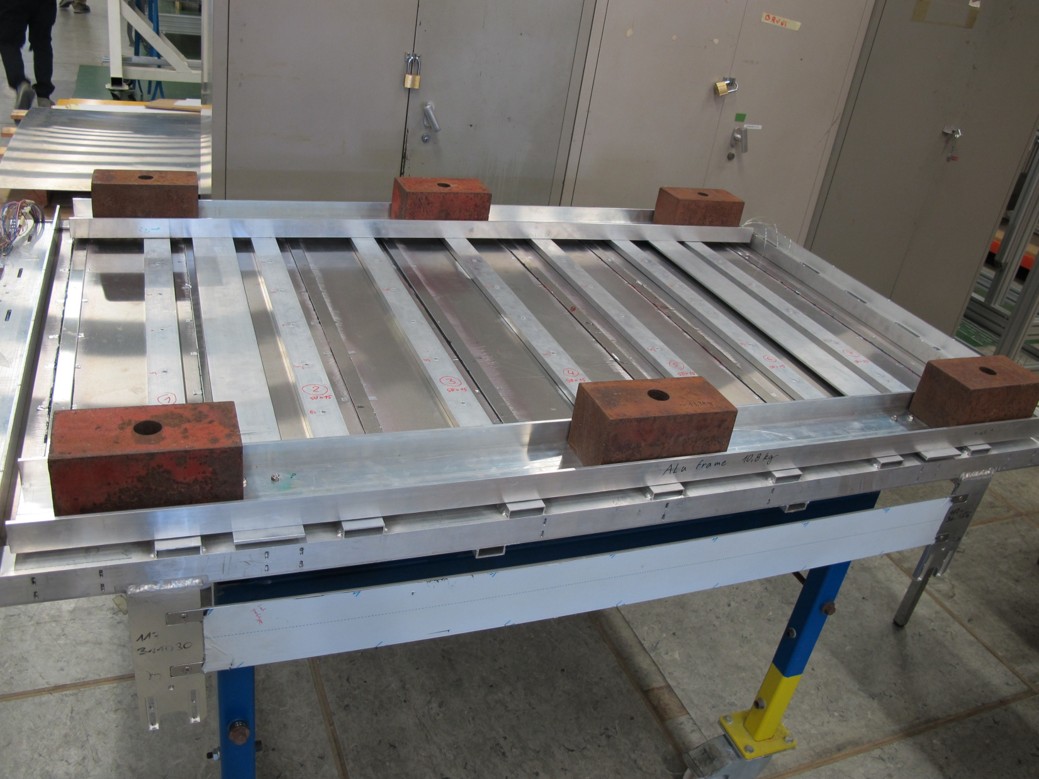}
    \caption{Upper frame installation through weights}
    \label{fig:weight}
\end{subfigure}
\hfill
\begin{subfigure}{0.45\textwidth}
    \centering
    \includegraphics[width=\textwidth]{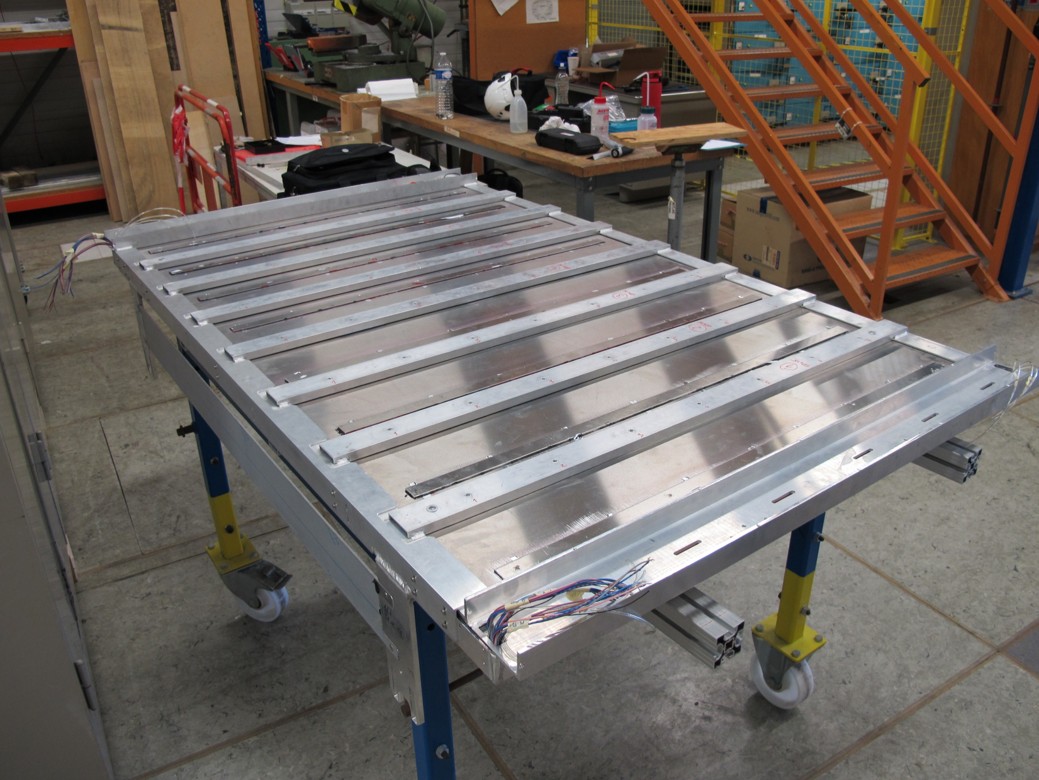}
    \caption{Assembled triplet}
    \label{fig:tripfin}
\end{subfigure}

\caption{Installation of the mechanical frame\label{fig:instmech}}

\end{figure}

Once all singlets are in position, the process of closing the mechanical structure begins, as illustrated in Figure \ref{fig:instmech}. The upper plate is carefully placed over the three singlets, ensuring that all services pass correctly through the designated cut-outs. The upper frame is then closed using a set of six calibrated weights corresponding to a pressure of 4.3 mbar. This calibration was conducted by measuring the efficiency uniformity across the detector surface using a half-glued gas gap.\\After the mechanical structure is closed and the services are installed, the triplet is ready for the cosmic rays test.\

\section{Cosmic rays test stand}
\label{sec4}
The cosmic ray (CR) test stand area is organized into three stations: the assembly area, the test stand, and the infirmary. The assembly area is dedicated to assembling the triplets, as described in Section \ref{sec3}. The test stand is used for testing with cosmic rays, aiming to fully certify the triplet performance, including noise rate, efficiency, cluster size, and timing. The infirmary is a separate station equipped with minimal instrumentation to identify and troubleshoot any issues with problematic triplets. This setup allows the debugging phase to be decoupled from the actual testing, preventing bottlenecks that could cause delays in the production process. The mechanical structure of the test station, Data Acquisition System (DAQ), and testing procedures have been designed to fully characterize the detector and meet the ATLAS requirements, ensuring the delivery of 16 triplets per month.
\subsection{Cosmic rays test stand set-up}
\label{sec3-sub1}
\begin{figure}[!h]
\includegraphics[width=\textwidth]{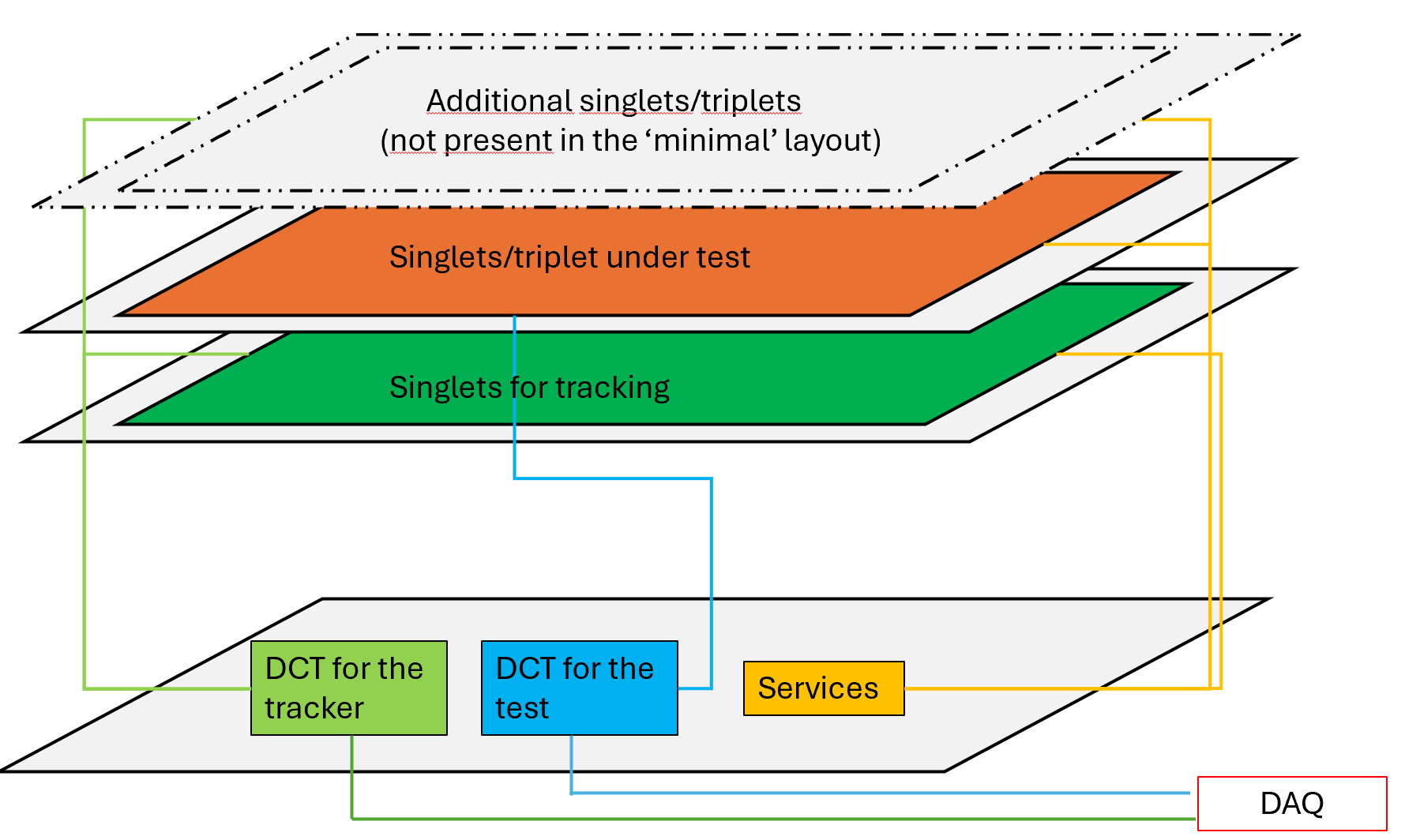}
\caption{Scheme of the CR set-up}\label{fig:CRscheme}
\end{figure}

The scheme of the cosmic rays set-up is shown in Figure \ref{fig:CRscheme}. The "minimal layout" consists in a movable table with three layers and made of Bosch profiles. 

The bottom layer, located near the wheels and separated from the other two, houses all the test and chamber services, including Data Transmission and Collection (DCT), the low voltage system, and a laptop for analysis. This layout ensures easy installation and quick test preparation. The middle layer hosts the external tracker, which consists of a spare triplet of RPCs (already certified and tested) with the same dimensions and layout as the triplet under test. This tracker also serves the function of monitoring. This layer is free to move, allowing it to be extracted from the table without affecting the other layers, facilitating maintenance and/or repairs. The upper layer is at eye level and houses the triplet under test. It is placed as close as possible to the middle layer to ensure full acceptance during the test. The triplet under test and the trigger are aligned using bars that span both the middle and upper layers, within which the detectors are positioned. The Bosch profile layout allows for the easy installation of a fourth layer on top, where additional triplets can be tested using the same system. This possibility is limited by the number of available DCT channels, which is still unknown. Since the performance of the detector depends on environmental parameters and the quality of the gas, sensors for temperature, pressure, and humidity, along with the gas flow, will be installed. These parameters will be continuously monitored and recorded by an external system (DCS), along with the detector HV and current.

\subsection{DAQ system}
\label{sec3-sub2}
\begin{figure}[!h]
\includegraphics[width=\textwidth]{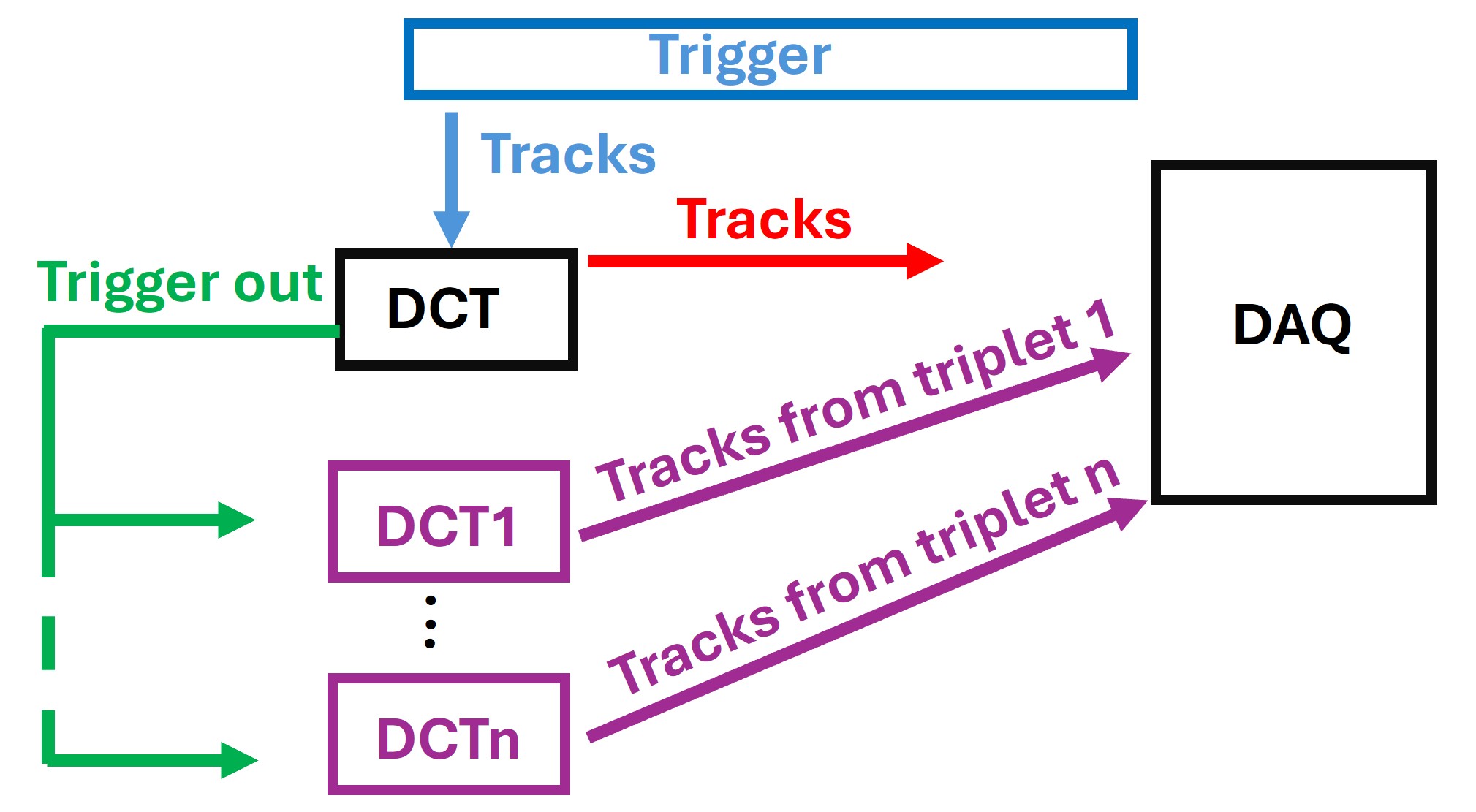}
\caption{Scheme of the DAQ}\label{fig:daq}
\end{figure}

The DAQ system, shown in Figure \ref{fig:daq}, consists of DCTs dedicated to the acquisition of both the monitor and the triplet(s) under test, as well as an acquisition laptop for offline analysis.\\One DCT will capture both tracks data and trigger-out coincidences from the monitor chamber, depending on the ongoing test, as discussed in Section \ref{sec3-sub3}. The track data are sent to the acquisition system, while the trigger-out signal is sent to the other DCT(s) as the input trigger. These DCT(s) collect the hits from the triplet(s) under test for the triggered event and transmit the data to the acquisition system for offline analysis. This system ensures the independence between the trigger and the triplet under test, as well as among multiple triplets under test, providing robustness and facilitating easy debugging.

\subsection{Triplet certification tests}
\label{sec3-sub3}
The layout of the cosmic rays test stand allows to perform all the tests needed in order to fully characterize the triplet under test before the installation in  the ATLAS cavern and fastly debug the system. There are three types of test foreseen for the certification of the triplet:

\begin{itemize}
    \item Test with random trigger, aimed at studying the detector and electronics noise.
    \item Test with external tracker, aimed at characterizing and certifying the triplet performance, including noise, efficiency, cluster size, and timing.
    \item Test with autotrigger, aimed at studying correlated noise among singlets within the same triplet (so-called self-induced noise).
\end{itemize}
The tests have been selected in order to study the main RPC characteristic as similar as possible to the final layout, minimizing potential issues during and after the installation in the cavern.
\subsubsection{Tests with random trigger}
\label{sec3-sub3-sub1}
\begin{figure}[!htbp]
\centering
\begin{subfigure}{0.45\textwidth} 
    \centering
    \includegraphics[width=\textwidth]{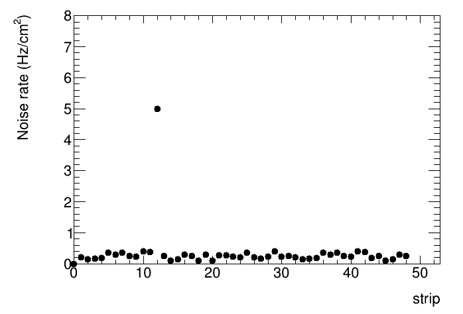}
    \caption{}
    \label{fig:nrbad}
\end{subfigure}
\hfill 
\begin{subfigure}{0.45\textwidth}
    \centering
    \includegraphics[width=\textwidth]{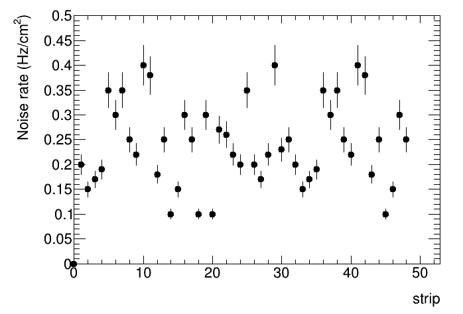}
    \caption{}
    \label{fig:nrgood}
\end{subfigure}
\caption{Detector noise evaluated at 5.7 kV using a random trigger, for a case with a noisy strip (a) and for a detector without any noisy strips (b). \label{fig:nr}}
\end{figure}

The test using the random trigger is aimed to the study of the detector and electronics noise and is performed in three different configurations:
\begin{itemize}
    \item The RPCs have the HV turned off and the LV turned on. As a result, the only noise observed is due to the FE electronics, allowing the study of the electronics noise.
    \item The RPCs are set to a fixed voltage of 4 kV, where the detector is "on" from the HV perspective but "off" in terms of gas ionization. This configuration allows for the identification of any noise originating from the HV.
    \item The RPCs are set to a fixed voltage corresponding to the Working Point (WP) voltage. At this stage, the detector is efficient, and since all electronics and HV noise have been excluded, any remaining noise can only be attributed to the detector itself.
\end{itemize}
Figure \ref{fig:nr} illustrates an example of noise evaluation for a detector fixed at 5.7 kV, obtained using a prototype. Figure \ref{fig:nrbad} shows a case where the acceptance limit for the ATLAS noise rate requirement of below 1 Hz/$\rm{cm^{2}}$ is not met. In such cases, the noise source is identified and addressed to bring the noise rate below the required level, as demonstrated in Figure \ref{fig:nrgood}.
\renewcommand\thesubfigure{\alph{subfigure}} 

\subsubsection{Test with cosmic rays}
\label{sec3-sub3-sub2}
\begin{figure}[!h]
\includegraphics[width=\textwidth]{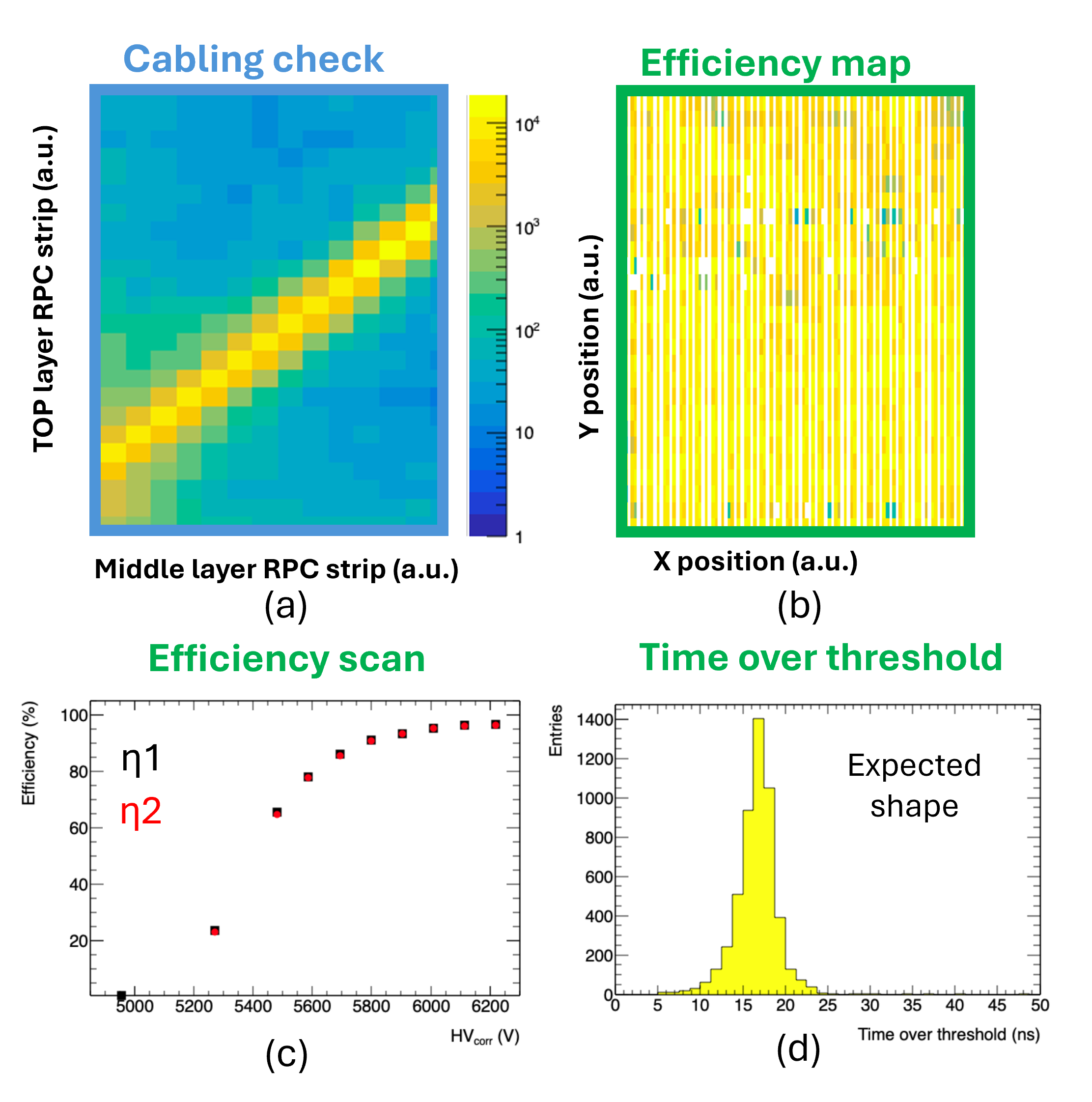}
\caption{Cabling check (a), Efficiency map (b), efficiency as a function of the high voltage (c) and time over threshold (d)}\label{fig:rpcperf}
\end{figure}

The test with cosmic rays is performed using the external tracker as a trigger, as mentioned in Section \ref{sec3-sub3}. This test is performed in three steps. The first step involves cabling the chamber and conducting a related control check, as shown in Figure \ref{fig:rpcperf}a. During this phase, the signal cables are connected to the RPC, and a quick scan at the working point is performed to verify that the cabling has been done correctly. A 2D plot of the hits is generated, and if the cabling is correct, the result will show a diagonal pattern, as illustrated in the figure.\\The second phase involves the so-called "independence test." In this configuration, one RPC is set to the working point voltage, while the other two remain off with their electronics switched on. This test, designed to detect any correlated noise among the singlets, is repeated for each possible permutation. Once the previous phases are successfully completed, an efficiency scan at high statistics will be performed, along with a performance study at the working point.
Figures \ref{fig:rpcperf}b, c, and d illustrate some of the main RPC parameters that will be evaluated during this phase. 
Figure \ref{fig:rpcperf}b shows the efficiency at the working point across the entire RPC area, using two out of three singlets as monitors in combination with the tracker. This setup provides the segmentation and correct normalization factor for the efficiency map. This test ensures that the RPCs achieve their intrinsic efficiency of 98.6\% across the entire area and verifies that the detector is undamaged. Figure \ref{fig:rpcperf}c illustrates the RPC efficiency as a function of the high voltage, measured using both panels of strips. This test enables the study of the efficiency curve and ensures that the detector average behavior aligns with expectations. One of the main RPC parameters, the time over threshold, is shown in Figure \ref{fig:rpcperf}d. This distribution is crucial as it is proportional to the electronic charge distribution, up to a multiplicative factor. The expected distribution is peaked around a mean value and confined within a narrow range, without any tails. The presence of tails would indicate high-charge events, such as streamers.
\subsubsection{Test in autotrigger mode}
\label{sec3-sub3-sub3}
The final set of measurements is conducted in autotrigger mode, as the BI RPC will function as a trigger chamber in the ATLAS experiment. This test is performed with one RPC off and the other two on, and is designed to study any fake muons caused by self-induced correlated noise. This configuration is the only one that allows the study of fake muons, as there are no discriminating variables to identify these events with an external trigger or when all detectors are switched on. Fake muons are consistent with the expected behavior of a muon signal, both in terms of timing (within 1 ns) and spatial position (same $\rm{\eta_{1}}$-$\rm{\eta_{2}}$ strip). If such events occur, they are not uniformly distributed among singlets and/or triplets and could lead to a potential loss in efficiency.
\section{Integration with s-MDT}
\label{secintegr}

\begin{figure}[!htbp]
\centering
\begin{subfigure}{0.45\textwidth} 
    \centering
    \includegraphics[width=\textwidth]{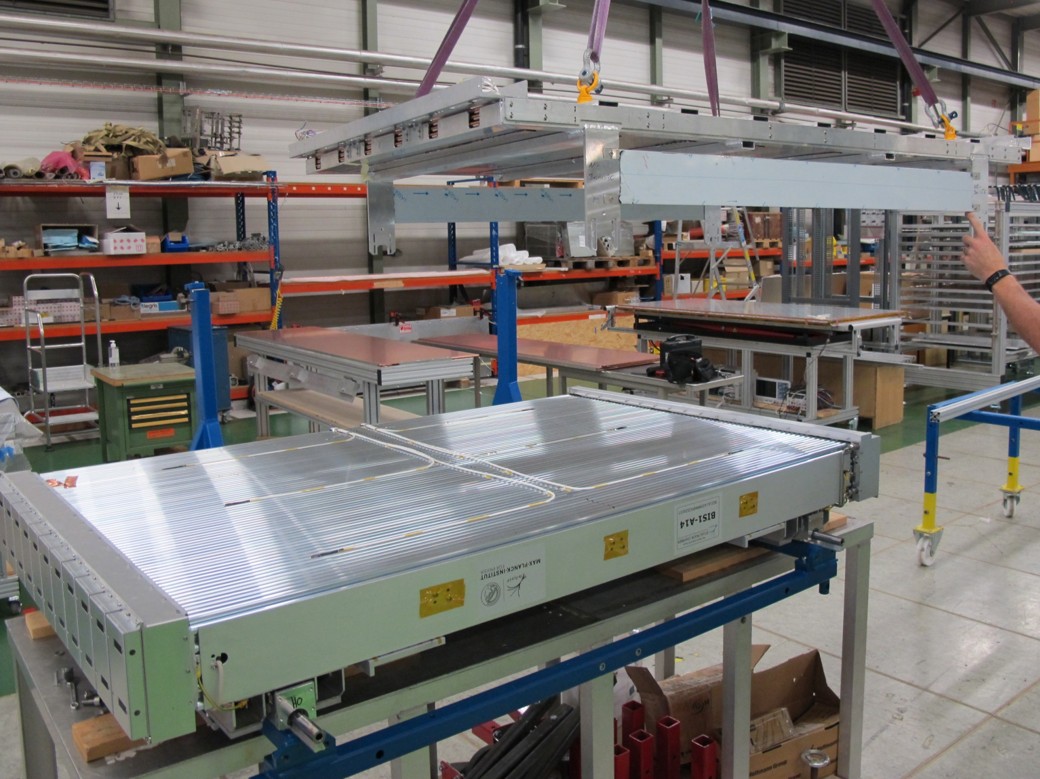}
    \caption{}
    \label{fig:int1}
\end{subfigure}
\hfill 
\begin{subfigure}{0.45\textwidth}
    \centering
    \includegraphics[width=\textwidth]{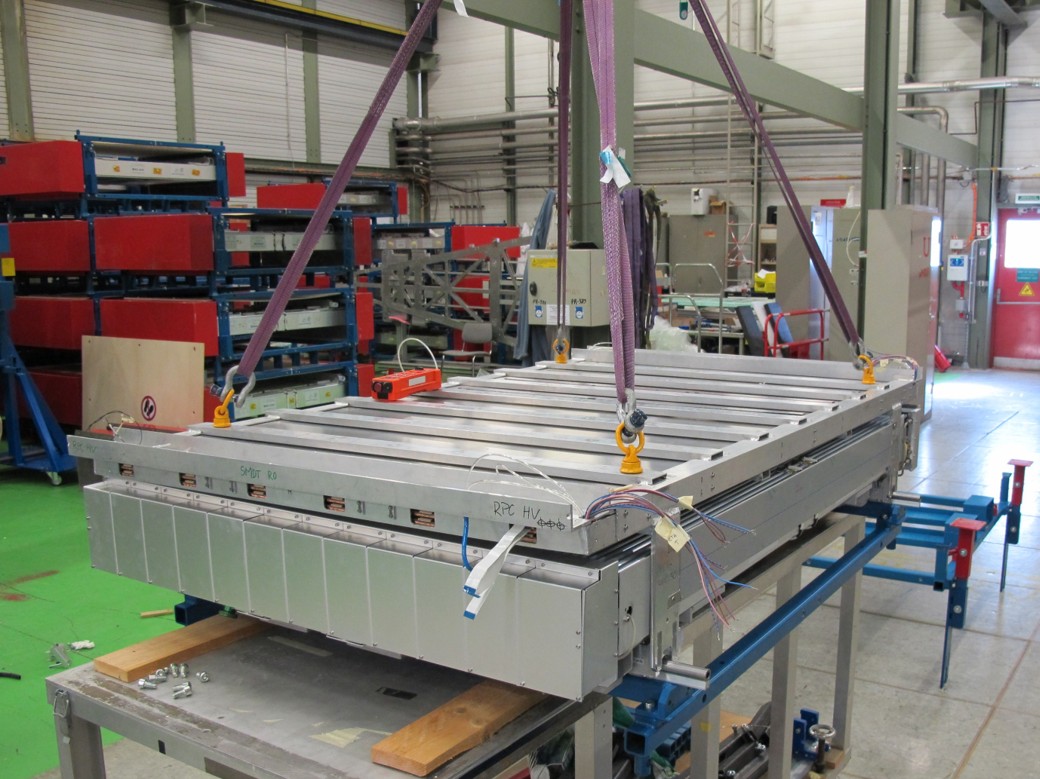}
    \caption{}
    \label{fig:int2}
\end{subfigure}

\vspace{0.5cm} 

\begin{subfigure}{0.45\textwidth}
    \centering
    \includegraphics[width=\textwidth]{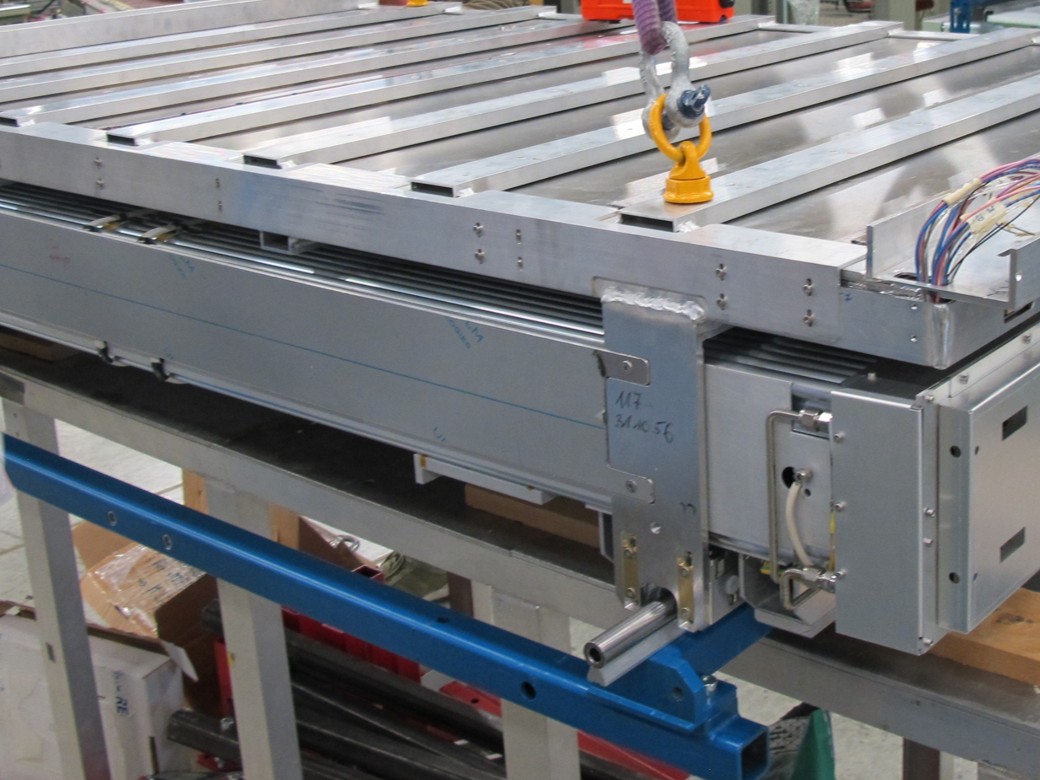}
    \caption{}
    \label{fig:int3}
\end{subfigure}
\hfill
\begin{subfigure}{0.45\textwidth}
    \centering
    \includegraphics[width=\textwidth]{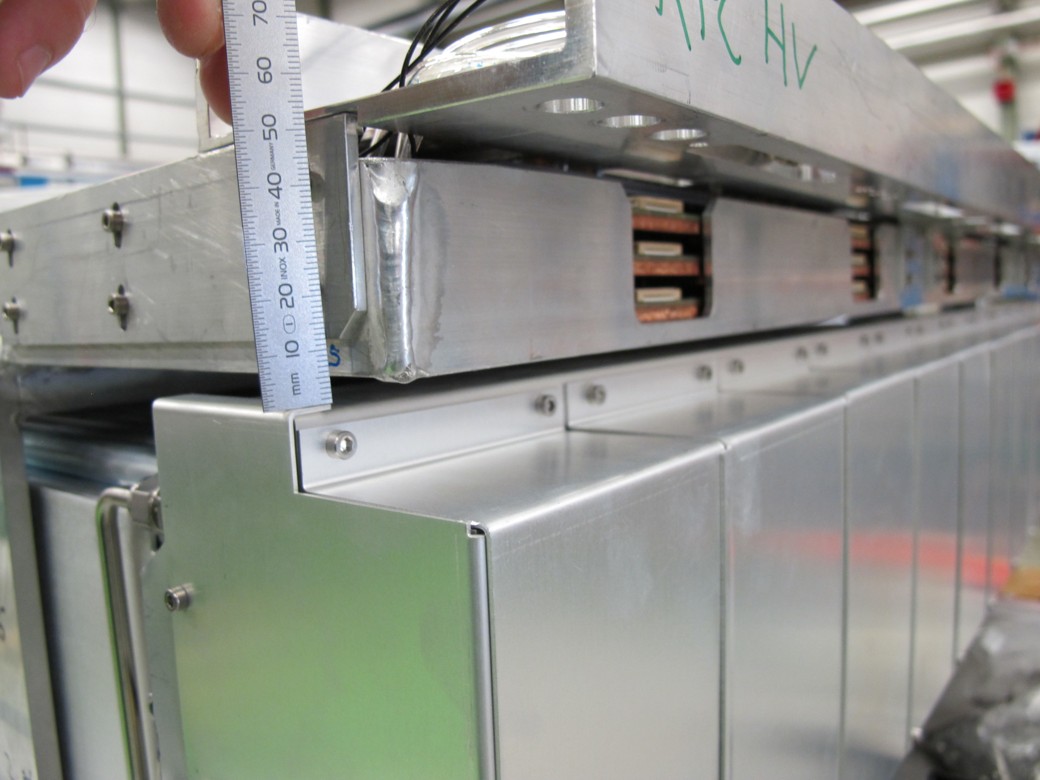}
    \caption{}
    \label{fig:int4}
\end{subfigure}

\caption{Integration of the RPC with s-MDT\label{fig:intsmdt}}

\end{figure}
The detectors that pass the tests discussed in Section \ref{sec3-sub3-sub2} are shipped to CERN, where they undergo the same tests to ensure that no damage occurred during transport. Once the detector is fully certified, it is integrated with the s-MDT following the steps outlined in Figure \ref{fig:intsmdt}.
The integration process involves placing the RPC chamber onto the s-MDT detector and installing the sliding plates onto the RPC frame feet. Next, the positions of the four RPC frame feet are set and fixed, and the RPC position is verified (envelope check). Finally, the gaps between the RPC and s-MDT chambers are checked to ensure there is no mechanical contact between the two. After integration, both detectors undergo the "independence test" to demonstrate electrical independence. This test consists of switching on the RPC and test the s-MDT while it is off, and vice versa, to check for any signal detection. Once these tests are passed, the BIS chamber is fully certified and ready for installation in the ATLAS experiment. 
\newpage
\section{Conclusions}
The RPC triplet integration project within the ATLAS detector has achieved significant milestones in the testing and certification phases. Testing methods have been developed and refined to ensure the detectors meet the required performance criteria, including efficiency, noise, and timing, guaranteeing high quality and reliability for both individual detectors and the entire triplet system. Cosmic ray tests, performed on each individual detector and subsequently on the full triplet, have been validated to be consistent with the conditions expected in the actual experiment.\\
Furthermore, the procedure of integration of the detectors with the s-MDT was successfully completed, in order to ensure both electrical and mechanical independence between the two systems, and to confirm that the detectors can operate optimally in the complex environment of the ATLAS experiment. Independence tests, along with precise mechanical tolerance and alignment checks, have been scheduled to confirm the functionality of the system before the installation.\\
The work will lay the foundation for the installation of the RPC triplets into the ATLAS detector, enabling the achievement of the monthly production and testing targets. With these tests successfully completed and the detectors fully certified, the RPC triplets will be ready to contribute to the success of the ATLAS experiment.








\end{document}